\documentclass{article}

 \usepackage[preprint]{neurips_2026}

\usepackage{xcolor}
\usepackage{listings}

\definecolor{codegray}{gray}{0.45}
\definecolor{codebg}{RGB}{248,248,248}
\definecolor{codeblue}{RGB}{0,70,140}
\definecolor{codegreen}{RGB}{0,110,60}

\lstdefinestyle{pythonappendix}{
    language=Python,
    basicstyle=\ttfamily\scriptsize,
    keywordstyle=\color{codeblue}\bfseries,
    commentstyle=\color{codegreen},
    stringstyle=\color{purple},
    numbers=left,
    numberstyle=\tiny\color{codegray},
    stepnumber=1,
    numbersep=6pt,
    backgroundcolor=\color{codebg},
    frame=single,
    rulecolor=\color{black!30},
    breaklines=true,
    breakatwhitespace=false,
    columns=fullflexible,
    keepspaces=true,
    showstringspaces=false,
    tabsize=4,
    captionpos=b
}

\usepackage{amsmath}

\usepackage[utf8]{inputenc} 
\usepackage[T1]{fontenc}    
\usepackage{hyperref}       
\usepackage{url}            
\usepackage{booktabs}       
\usepackage{amsfonts}       
\usepackage{nicefrac}       
\usepackage{microtype}      
\usepackage{xcolor}         
 \usepackage{booktabs}
 \usepackage{booktabs,xcolor,multirow,rotating}
\usepackage{cleveref}

\usepackage{titlesec}
\titlespacing*{\paragraph}{0pt}{0pt}{1em}

 \definecolor{myred}{HTML}{F54254}
\definecolor{myorange}{HTML}{FFB135}
\definecolor{mygreen}{HTML}{10BD35}
\definecolor{myblue}{HTML}{4e58cc}
\definecolor{myblue_light}{HTML}{7292da}
\definecolor{mypurple}{HTML}{9656ce}

\renewcommand{\phi}{\varphi}

\usepackage[table]{xcolor}
\usepackage{wrapfig}

\usepackage{hyperref}
\hypersetup{
    colorlinks=true,
    urlcolor=myblue,
    citecolor=myblue,
    linkcolor=myblue
}
\newcommand{\DATASET}{MPMWorlds}

\title{\DATASET: Material-Point-Method Simulations for Inferring and Extrapolating Physical Dynamics}

\author{%
  \v{Z}iga Kova\v{c}i\v{c} \\
  Cornell University \\
  \And
  Kevin Ellis \\
  Cornell University \\
}

\begin{document}

\maketitle

\begin{abstract}
To study the ability to infer physical dynamics from videos and extrapolate them forward in time, we assemble a dataset of 2D Material Point Method (MPM) physical simulations covering rich physical phenomena such as deformable objects, fluids, kinetic objects, and emitters.
We study code generation and video diffusion approaches on this dataset, identifying their strengths and weaknesses by varying the amount of physically relevant side information.
The code generation model, beyond giving a working demonstration of automatic synthesis of MPM simulations, reveals that such an approach struggles with inferring physical parameters from visual input, but relative to video diffusion, produces physically and temporally stable extrapolations forward in time, while the video diffusion model more strongly identifies geometric properties from visual input but produces physically implausible extrapolations. Project page: \href{https://zzigak.github.io/mpmworlds/}{https://zzigak.github.io/mpmworlds/}.
\end{abstract}

\section{Introduction}
Physical simulation is used across robotics, science and engineering, videogames, cognitive science, and the arts.
Up until recently, building a high-fidelity physical simulator handling rich physics, such as deformable objects, fluids, etc., meant painstaking programming using techniques such as Material Point Methods (MPM).
But could generative models of source code dispense with the need to hand-program new simulations---and could generative models of videos further dispense with the need to have any simulation code at all?
These questions are significant, because authoring precise simulation code is expensive and demands domain expertise, while generative video models might achieve much broader coverage of the everyday world than any symbolic simulation ever could.

The ability to infer the physical dynamics of a scene, and successfully extrapolate to future time points, is also a fundamental test of physical reasoning.
While the literature has investigated the ability of generative video models to extrapolate physics forward in time~\citep{kang2024far}, such works center on simple rigid body dynamics, not the richer dynamics such as fluids, deformable objects, and various different materials.

\begin{figure}
\includegraphics[width = \textwidth]{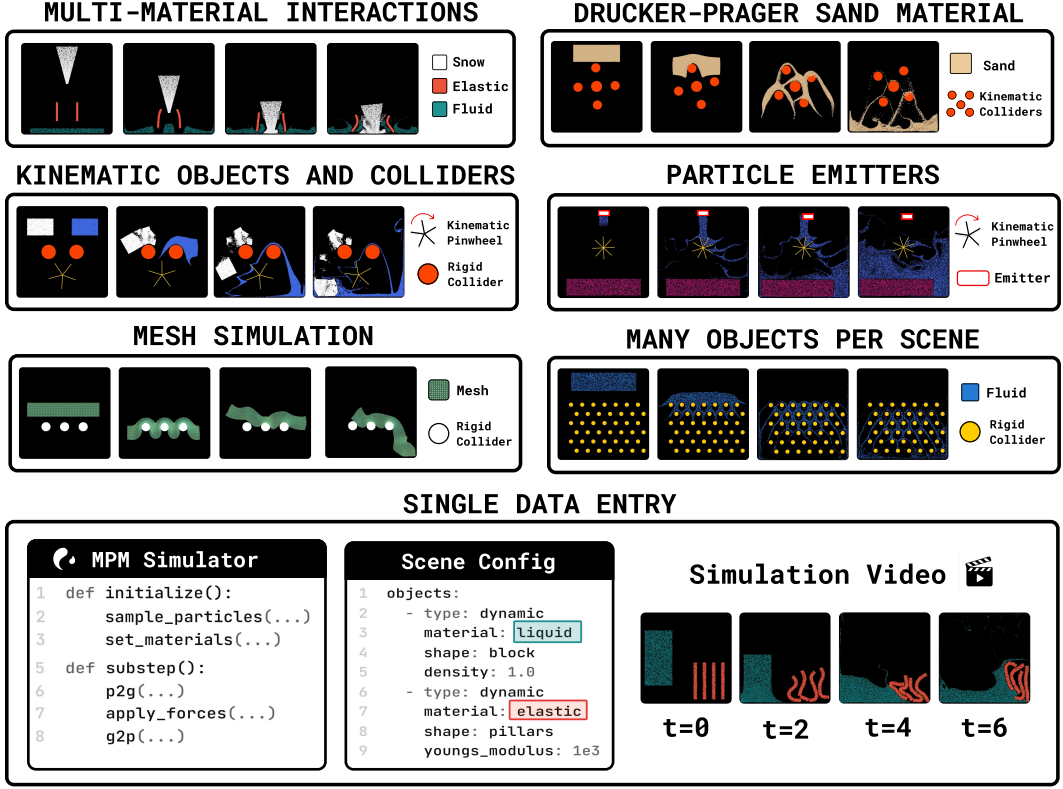}
\caption{\textbf{MPMWorlds dataset overview.} The dataset includes diverse materials and physical interactions. Each entry contains the simulator code, the scene configuration, and the resulting video. }
\label{fig:teaser}
\vspace{-10pt}
\end{figure}

To answer these questions, we assemble a new dataset, MPMWorld, of physical simulations  covering rich dynamics such as fluids, materials such as snow and sand, deformable bodies, and more (\Cref{fig:teaser}).
We then study the conditions under which generative models of source code (specifically VLMs) can produce accurate simulation code, given partial information such as a history of past frames, or privileged information such as underlying physical parameters.
We also study the conditions under which generative video models (specifically video diffusion models, `VDMs') can render the output of the desired simulation, without ever explicitly constructing the underlying code.

We find both methods achieve only partial success.
Code generation excels at consistent long-horizon dynamics,  but struggles with geometry.
Video diffusion excels at geometry but hallucinates dynamics and deteriorates over long time horizons.
We probe the models to understand which elements of the simulations they struggle with, identifying specific classes of materials that are surprisingly challenging (or easy) for each approach.

In total we contribute the following:
\vspace{-3pt}
\begin{enumerate}
    \item A dataset of 2D physical simulations covering fluids, deformable objects, rigid bodies, emitters, `motorized' objects such as pinwheels and conveyor belts, and more.
    Each simulation includes both \emph{source code}, \emph{scene configuration file}, and an associated \emph{video of the simulation running}.
\vspace{-3pt}
    \item Neural network models trained on this dataset.
    We train a variety of models which either generate physics simulation source code, or which generate raw frames, and which condition on a variety of information sources, such as past frames or privileged information such as the numerical values of physical parameters or initial positions. 
\vspace{-3pt}

    \item 
    The evaluative role of our dataset and models is to test learning-based methods for inferring physical dynamics, and extrapolating that inference forward in time.
    Our work analyzes which physical dynamics are challenging to reconstruct and extrapolate.
    But importantly the analysis offers constructive guidance for future work:
    By studying which forms of physically-relevant side information most improves the models, we hope to offer constructive guidance on what future modeling should focus on.
    
    
\end{enumerate}

\subsection{Problem Definition: Inferring and extrapolating physical dynamics.}
Given partial information about the physical dynamics of a scene, such as a short video clip, we take \emph{inferring physical dynamics} to mean the recovery (or reconstruction) of a forward simulation of that physical scene.
These problems are practically relevant because extrapolations forward in time can serve as a world model for e.g. planning or intuitive physics purposes, and inferring the underlying program is valuable for engineers or artists wishing to work with the physics code.
We consider a variety of partial information sources in part because end-users may wish to specify a physical simulation with more than just images or video, but also provide specific physical parameters, while keeping the rest unspecified.

The recovered physical simulation could be represented explicitly as source code, or it could be implicitly represented as neural activations, but in either case, it should support \emph{extrapolation to future time points}, meaning that we can predict what the scene will look like at any point in the future.

We consider scenes generated by physics simulation programs.
We  write $\rho$ for one such program; different programs simulate different scenes with different objects, materials, etc.
A program $\rho$ produces a video $v$ comprising a sequence of frames.
Given a prefix of $v$ up to time $t$, written $v_{\leq t}$, inferring the physical dynamics as a program means to recover a $\hat\rho$ equivalent to $\rho$.
Extrapolating to future time points means predicting a video $\hat v = \hat v_{\leq t} || \hat v_{>t}$ (given 
$v_{\leq t}$).

\begin{figure}
\includegraphics[width = \textwidth]{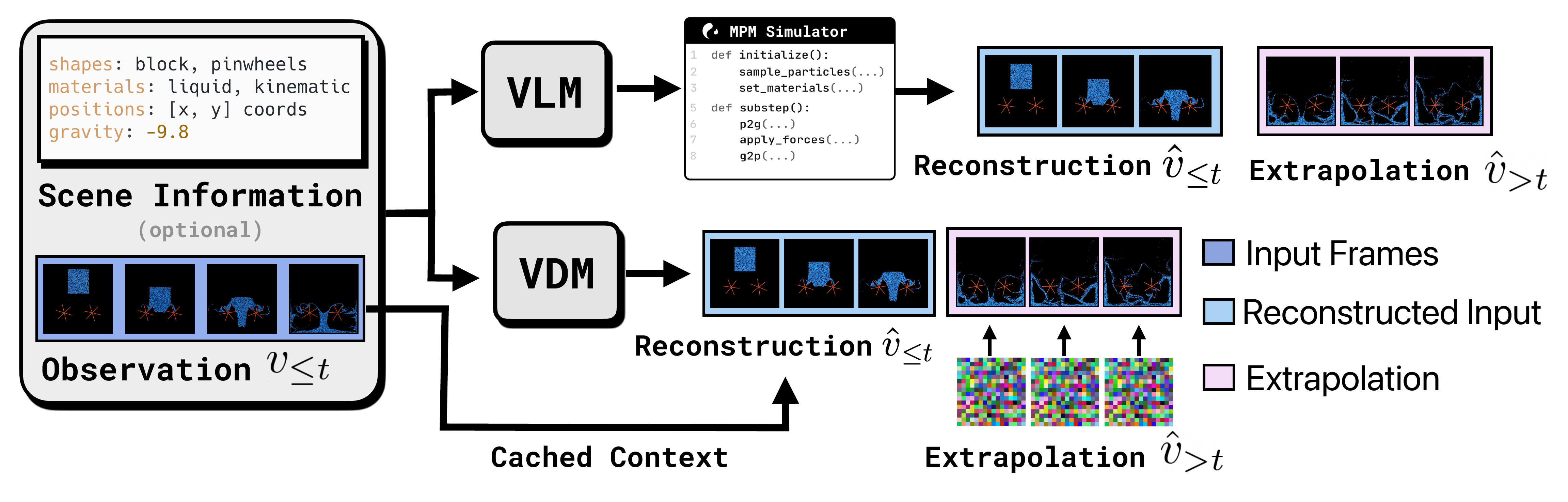}
\caption{\textbf{Reconstruction and extrapolation pipelines.} Models must predict a full video sequence ($\hat{v} = \hat{v}_{\le t} \parallel \hat{v}_{>t}$) from an initial observation ($v_{\le t}$). \textbf{Top:} The VLM synthesizes a simulation program that executes from $t=0$, explicitly reconstructing the input (blue) alongside the future extrapolation (pink). \textbf{Bottom:} The VDM operates in pixel space, using the input strictly as conditioning to generate future frames.}
\label{fig:task}
\vspace{-10pt}
\end{figure}

Given the complexity of physical dynamics that we study, we do not expect inference and extrapolation to be easy.
To understand why a model struggles with inference or extrapolation for a particular scene, we can provide physically-relevant \emph{side information}, such as the positions of objects, their densities, etc.
We extract different forms of side information from the underlying program, written
$f(\rho)$.
Models predict either a video or a program,  i.e. estimate either $p(v\mid v_{\leq t},f(\rho))$ or $p(\rho\mid v_{\leq t},f(\rho))$. Figure ~\ref{fig:task} illustrates the two inference-and-extrapolation pipelines studied in this work: executable simulation synthesis via VLMs and direct pixel-space continuation via VDMs.



\section{Related Works}

\paragraph{Intuitive Physics and Physical Reasoning.}
Prior work studies physical reasoning as the ability to infer and predict physical dynamics from visual observations. Cognitive theories propose that human intuitive physics operates through approximate internal simulation \citep{ullman2017mind}, while benchmarks such as Physion \citep{bear2021physion} and Physion++ \citep{tung2023physion++} evaluate physical understanding in controlled environments. Learned relational and object-centric dynamics models include Interaction Networks \citep{battaglia2016interaction}, latent physical-property inference from videos \citep{wu2015galileo,wu2016physics}, and program-like scene representations \citep{liu2019learning}. Our work instead studies whether modern generative models can recover executable or implicit forward models of complex physical scenes.

\paragraph{Programmatic Physical Reasoning and Simulation Code Generation.}
Recent work explores executable or symbolic representations for physical reasoning. Scene2Prog \citep{liu2019learning} infers programmatic structure from visual scenes, VisPhyWorld \citep{liang2026visphyworldprobingphysicalreasoning} studies code-driven physical reconstruction, and PhysCodeBench \citep{xie2026physcodebench} evaluates LLM-based physics simulation synthesis. Our setting instead couples executable MPM programs with rendered rollouts and structured scene configurations, enabling controlled comparison between code-space and pixel-space prediction for deformable multi-material dynamics.

\paragraph{Video Generation Models as World Models.}


Recent work increasingly views generative video models as implicit world models \citep{ha2018world,bruce2024genie}. Several benchmarks evaluate physical plausibility in generated videos, including VideoPhy \citep{bansal2024videophy,bansal2025videophy}, LikePhys \citep{yuan2025likephys}, PhyWorldBench \citep{gu2025phyworldbench}, and analyses of physical-law consistency in video generation \citep{kang2024far}. Unlike prior work focused primarily on perceptual plausibility or short-horizon events \citep{qiu2025phybench,meng2024towards}, our setting studies long-horizon continuation of deformable multi-material dynamics with paired executable simulations and structured scene representations.

\paragraph{Physical Reasoning Benchmarks.}
Prior benchmarks evaluate physical reasoning through mechanics puzzles, causal reasoning, and interactive simulation, including PHYRE \citep{bakhtin2019phyre}, CLEVRER \citep{yi2019clevrer}, Physion \citep{bear2021physion}, Physion++ \citep{tung2023physion++}, and I-PHYRE \citep{li2023phyre}. PHYBench \citep{qiu2025phybench} studies formal physics reasoning in language models. In contrast, our dataset evaluates generative physical continuation, where models must extrapolate future dynamics either through executable simulation synthesis or direct video generation.

\section{MPMWorlds Dataset}


To study physical dynamics inference and extrapolation, we introduce \textbf{\DATASET}, a simulation-based dataset of physical scenes. Each instance includes (i) executable simulation code, (ii) a structured configuration specifying the scene, and (iii) a rendered video of the resulting dynamics. The scene configuration specifies object types, materials, geometry, initial conditions, and appearance, while the simulation code implements the corresponding physical process. This paired representation enables controlled study of dynamics extrapolation, physical state inference, sensitivity to input modalities, and the role of explicit intermediate representations such as code.

Each configuration is a structured YAML file containing object entries with fields for object type, geometry, material model, initial state, appearance, and optional kinematic or emitter behavior. For example, a dynamic body specifies its shape, color, constitutive model, material parameters, and initial position and velocity; emitters specify spawn region, material, and emission rate.

\paragraph{Dataset Characteristics.}
We implement the simulations in 2D using the \textbf{Material Point Method} (MPM), rendering videos at $512 \times 512$ resolution and 30 FPS for 10 seconds. All simulations are written in Taichi, with each program explicitly defining the MPM state update, scene initialization, physical parameters, and time-stepping loop, without relying on high-level simulation APIs.

We use MPM due to its stability and its ability to model a wide range of materials and interactions within a unified framework. It is also widely adopted across computer graphics, visual effects, and engineering communities, making it a practical and well-established choice for simulating diverse physical phenomena. The dataset includes diverse dynamic scenes with deformable and multi-material interactions, including liquids, elastic (particle-based and FEM-based), plastic, snow, sand (including wet interactions), viscoplastic and viscoelastic materials, as well as rigid and kinematic objects \ref{fig:stats}. We also include emitters that continuously introduce material into the scene (e.g., fountains or rain). All scenes are rendered on a high-contrast background to minimize confounding visual factors such as texture or lighting.


\vspace{-5pt}
\subsection{Dataset Creation Pipeline}

The dataset is generated using a partially synthetic pipeline combining human-written programs, LLM-based synthetic generation, and data augmentation through small perturbations. It is created entirely from scratch and does not rely on existing datasets or sources. The dataset generation pipeline has a hierarchical structure and we explain the steps below:

\paragraph{Step 1: Seed Simulations.}
We manually write $35$ diverse ``seed'' simulations with varying object layouts, materials, geometries, initial velocities, kinematic elements (e.g., pistons, gates, rotating objects), and rigid colliders. Simulations are paired with a structured scene YAML config.


\paragraph{Step 2: Base Simulation Generation and Validation.}
We expand the dataset in the style of Self-Instruct~\citep{wang2023self} by prompting large language models (ChatGPT o3 and Gemini 2.5 Pro) with $2$--$3$ seed examples to generate new simulation programs and corresponding scene configurations, by combining and extending the provided human written code. Generated simulations are executed and filtered using automated checks for runtime failures, trivial outputs, and severe numerical instability. We additionally validate consistency between generated programs and configuration files using an LLM-based verification pass checking object counts, materials, and initial conditions, with flagged failures manually inspected. This process yields approximately $1$k valid \textbf{Parent} simulations.

We further expand the dataset by prompting the LLM to modify previously generated scenes through changes such as adding or removing objects, altering materials or geometries, and modifying interactions, resulting in approximately 9k \textbf{Child} simulations. Together, the Parent and Child simulations constitute the core ~10k \textbf{Scene Templates} that define the unique physical setups in our dataset.

\paragraph{Step 3: Automated Perturbations.}
We augment the dataset by perturbing numerical parameters, e.g. object positions, subject to spatial constraints that avoid invalid layouts, such as overlapping objects or objects outside the domain. This reduces bias from LLM-chosen initial configurations, introduces variability in initial conditions, and aids generalization.


\subsection{Dataset Statistics}
\textbf{\DATASET} contains 95,805 rendered simulations, including 9,204 Scene Templates (comprising the full set of Parent and Child simulations) and associated perturbations. Base scenes define the physical setup, while perturbations vary numerical parameters.

\begin{figure}
\includegraphics[width = \textwidth]{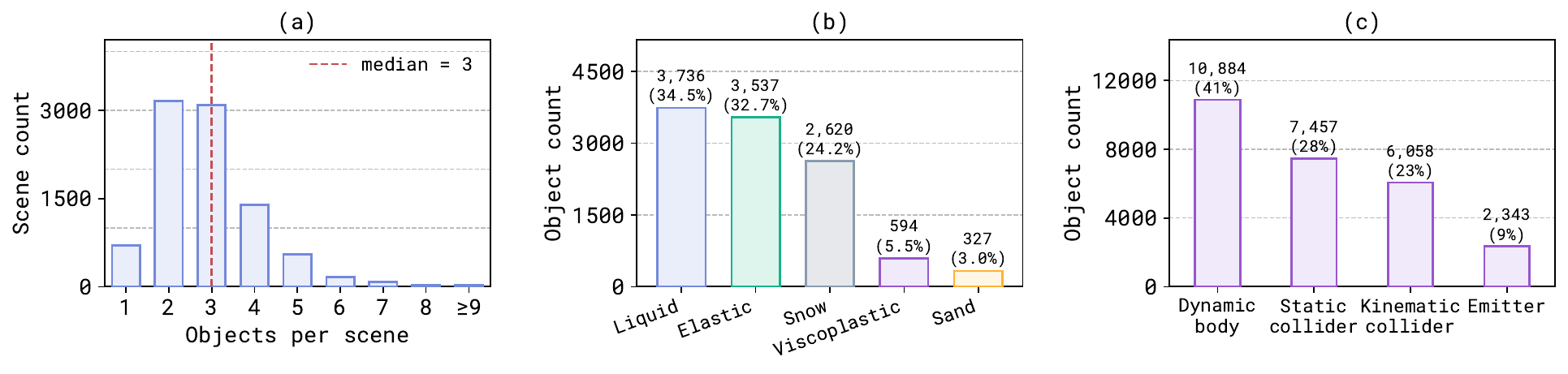}
\caption{Dataset statistics for \textbf{\DATASET} across base scenes. (a) Distribution of number of object per scene. (b) Distribution of dynamic-body material types across the dataset. (c) Distribution of object types illustrating the diversity of interaction mechanisms present in the dataset.}
\label{fig:stats}
\end{figure}

Figure~\ref{fig:stats} summarizes dataset statistics. Scenes have a median of 3 objects and mean of 2.9 objects, with a tail of more complex interactions involving dynamic bodies, colliders, and emitters. The dataset spans multiple classes of physical materials, including liquids, elastic materials, snow, viscoplastic materials, and sand; liquids and elastic materials are most common, followed by snow, while sand and viscoplastic materials provide additional deformation regimes. Overall, 22.5\% of scenes contain at least two distinct material types, and 5.3\% contain three or more. 

Object types include dynamic deformable bodies, static colliders, kinematic colliders, and emitters, enabling interactions such as moving boundaries, externally driven motion, and continuous material injection, substantially increasing the diversity and difficulty of long-horizon dynamics prediction.

\section{Experimental Setup and Evaluation}

\subsection{Input Conditions}
We evaluate each model under different input regimes designed to identify which physical properties prove  difficult to infer from visual observations alone and if one type of information is much easier to infer from visual information for one approach (VLM or VDM) than the other.

\paragraph{Frames Only.}
Models input only a video prefix of approximately 2.5 seconds ($v_{\leq t}$, where $t=2.5\text{sec}$). This is the minimal-information or vision-only setting, so the models have to infer all the information from vision.

\paragraph{Full Configuration.}
Models receive the video prefix $v_{\leq t}$ together with the complete scene configuration, including object positions, material types, geometry, initial conditions, and physical parameters. This is the maximal-information setting:
$f(\rho)$ gives the whole YAML file.

\paragraph{No Materials.}
Models receive the video prefix $v_{\leq t}$ and the scene configuration with material identities and material-specific parameters removed. This tests inference of material properties from visual evidence alone.
Here $f(\rho)$ returns the YAML but with material properties removed.

\paragraph{No Positions.}
Models receive the video prefix $v_{\leq t}$ and the scene configuration with explicit positional information removed. This tests whether models can infer object layout and initial state from the observed frames.
Here $f(\rho)$ returns the YAML but with initial positions removed.

\subsection{Models}
We compare two types of models, a code generation model (VLM) which produces executable simulation code against a video diffusion model (VDM) that directly predicts future frames.

\paragraph{Vision-Language Models.}
VLMs predict future dynamics indirectly by generating executable MPM simulation code. Given prefix frames and optional configuration information, the VLM outputs a full Python (Taichi) simulation program $\hat\rho$, which when executed predicts video $\hat{v}$.

We use Qwen2.5-VL-7B-Instruct as the base VLM. At inference time, the model receives sampled frames from the prefix, optional side information, and a lightweight program scaffold containing only boilerplate imports and function signatures. The scaffold fixes non-semantic code structure, such as function names and ordering, while the model generates the simulation logic, object initialization, material parameters, and update rules. This focuses learning on the physically meaningful parts of the program rather than arbitrary formatting choices.

\paragraph{Video Diffusion Models.}
VDMs predict future dynamics directly in pixel space without constructing an explicit physical representation. We use the 14B LongCat video continuation model as our diffusion baseline. The model is conditioned on the observed video prefix and, when available, the same textual scene information used for the VLM input regimes.

\subsection{Training and Inference}
For all experiments, we construct a 10k-example subset of the dataset with increased sampling weight on base scenes. We evaluate two held-out regimes enabled by the hierarchical dataset structure. The primary \textbf{test split} is partitioned at the Parent simulation level: all modifications (Children) and perturbations derived from a held-out base simulation are assigned to test, so models never observe the underlying scene template during training. This evaluates generalization to entirely novel physical scenes.

We additionally construct a \textbf{modification validation split} by withholding some modified descendants of training base scenes. This split intentionally preserves partial template overlap while changing object layouts, materials, etc., providing an intermediate regime between exact memorization and fully novel scene generalization.

To improve training and inference efficiency, videos are resized to $256 \times 256$ resolution at 15 FPS. For VLM training, we use full supervised finetuning. For VDMs, we observed little difference between full finetuning and LoRA adaptation and therefore use LoRA for all diffusion experiments.

At inference time, we generate multiple candidate programs from the VLM and select among them using evaluation metrics computed on the predicted $\hat v_{\leq t}$. Specifically, we sample 15 candidate programs per test example. In contrast, we observe relatively low diversity across diffusion samples under different random seeds and therefore draw a single sample from the VDM. We use 50 denoising steps for diffusion inference, which empirically worked well and which calibrates the inference time compute budget of the two models:
50 denoising steps takes about as long as sampling 15 programs.

\subsection{Evaluation Metrics}
Evaluating long-horizon physical rollouts requires metrics that capture not only pixel-level similarity, but also temporal consistency, motion fidelity, appearance preservation, and catastrophic physical failures such as collapse or disappearance.

Unlike classical simulation benchmarks, our setting does not provide direct access to particle trajectories or physical state during evaluation. Additionally, many scenes contain highly deformable materials whose shape and topology evolve over time, making standard tracking- or correspondence-based metrics ill-defined.

Prior work on physical video generation primarily evaluates visual plausibility using  perceptual or VLM-judged metrics \citep{bansal2025videophy,yuan2025likephys,gu2025phyworldbench}. We evaluate five complementary aspects of physical continuation: \emph{spatial correspondence}, \emph{appearance and material composition}, \emph{motion fidelity}, \emph{internal physical stability}, and \emph{temporal coherence}. Full details about the metrics are provided in Appendix \ref{appendix:metrics}.

\paragraph{Mask IoU (mIoU).}
We measure spatial overlap between predicted and ground-truth foreground occupancy masks. For each frame, RGB images are thresholded into binary foreground masks and the Intersection-over-Union is computed frame-wise.  

\paragraph{Object Collapse Score (OCS).}
We measure prediction-internal temporal stability by checking whether foreground regions generated early in the predicted extrapolation remain visible over time. We compute this by segmenting initial object/material regions, tracking their color-consistent visible area across frames, and measuring the maximum relative area loss. 

\paragraph{Windowed Motion Activity Error (W-MAE).}
We measure whether the prediction exhibits a similar amount of temporal activity as the ground-truth video, independent of exact spatial alignment. Motion activity is computed from frame-to-frame intensity changes aggregated over coarse spatial regions and temporally smoothed using a sliding window. W-MAE is the normalized difference between predicted and ground-truth activity magnitudes over time.

\paragraph{Color Total Variation (CTV).}
To evaluate appearance and material-composition accuracy, we compare foreground color distributions between predicted and ground-truth rollouts. At each frame, foreground pixels are discretized into RGB histograms and compared using total variation distance.

\paragraph{Temporal Anomaly Rate (RTSJ).}
We measure temporal coherence by detecting abrupt local appearance changes in predicted rollouts that exceed corresponding changes in the ground truth. We compute spatially aggregated temporal jump magnitudes from local color-histogram changes and flag a rollout as anomalous if its maximum excess jump exceeds a threshold. The reported anomaly rate is the fraction of test samples flagged as anomalous. Full details are provided in Appendix \ref{appendix:rtsj}.


\paragraph{Additional Metrics.}
We additionally report standard pixel-level metrics including MSE and sliced Earth Mover's Distance (SEMD) in supplementary experiments. However, we do not use them as primary evaluation metrics because they correlate poorly with long-horizon physical consistency.

\paragraph{Metrics enable Best-of-$N$ sampling for program generation.}
Because we sample multiple programs at test time, the VLM can benefit from Best-of-$N$ sampling:
Each program can be executed and its video compared to the ground truth prefix $v_{\leq t}$ under the above metrics, with the best-scoring program being returned.
Note that selection only reads the prefix $v_{\leq t}$; the suffix $v_{> t}$ remains hidden from the learner at test time and is used solely to evaluate the selected program.

\section{Results and Analysis}

Unless otherwise stated, results are on the \emph{fully held-out test split.}

\paragraph{How do executable simulation models compare to direct video prediction?}

\begin{figure}[t]
\includegraphics[width = \textwidth]{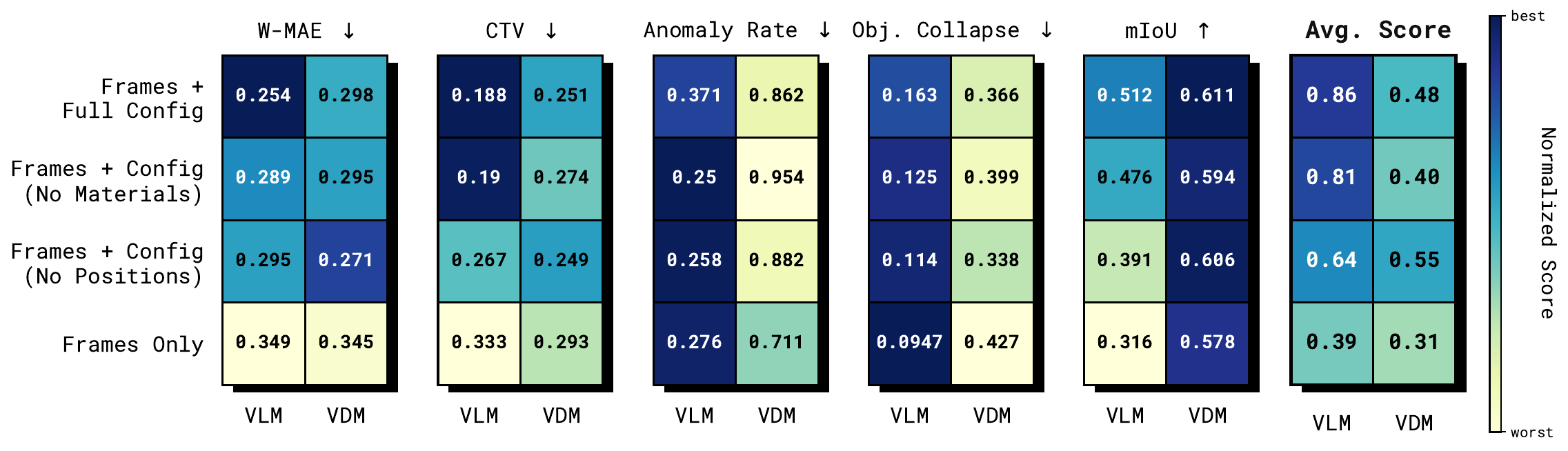}
\caption{Comparison of VLM- and VDM-based extrapolation across input conditions and evaluation metrics. Values are averaged over the held-out test split. Lower is better for W-MAE, CTV, anomaly rate, and object collapse, while higher is better for mIoU. The final column reports the mean normalized score across metrics for each model and input condition.}\label{fig:heatmap}
\vspace{-10pt}
\end{figure}

\begin{figure}[t]
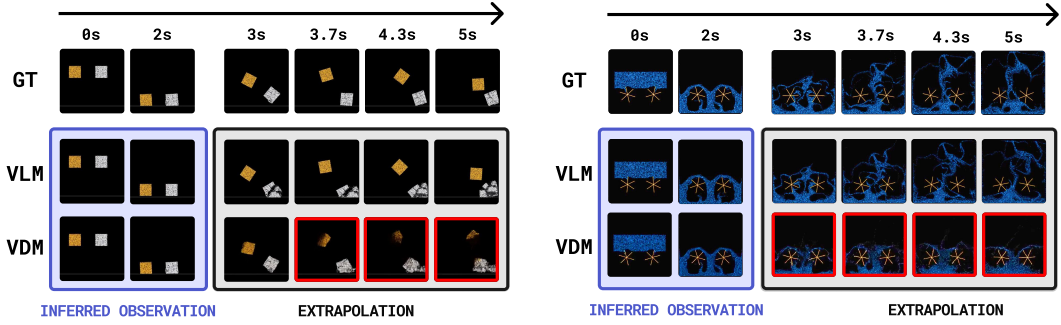

    \centering
    \begin{minipage}{0.48\textwidth}
        \centering
        \includegraphics[width=\linewidth]{images/anomaly_2.pdf}
    \end{minipage}\hfill
    \begin{minipage}{0.48\textwidth}
        \centering
        \includegraphics[width=\linewidth]{images/liquid_narrow.pdf}
    \end{minipage}
    \caption{\textbf{VDM failure modes in long-horizon physical extrapolation.} \textbf{Left:} In an elastoplastic scene, the VLM correctly maintains rigid object permanence and trajectory. The VDM suffers from object collapse, causing the bouncing block to fade and vanish during extrapolation. \textbf{Right:} Subjected to high-energy kinematic colliders, the VLM preserves complex fluid volume and splashing dynamics, whereas the VDM prediction fails to capture high frequency motion.}
    \label{fig:vdm_failures}
    \vspace{-10pt}
\end{figure}

Code generation and direct video prediction exhibit substantially different strengths and failure modes under long-horizon physical extrapolation task (Figure~\ref{fig:heatmap}, \Cref{tab:grouped_visual_metrics_top1}).
Code generation  (w/ VLMs) generally outperforms direct video prediction (w/ VDMs) on most metrics and input conditions, particularly on temporal stability and long-horizon consistency.
The largest differences appear in anomaly rate and object collapse, where executable simulations remain substantially more stable over time. Because generated programs are executed inside a physically grounded simulator, successful VLM generations typically preserve object persistence and temporal smoothness unless the predicted code itself becomes numerically unstable. In contrast, VDM predictions frequently exhibit hallucinated motion, disappearing regions, abrupt appearance changes, or temporally inconsistent dynamics, leading to significantly worse anomaly and collapse scores. Representative long-horizon failure cases are shown in Figure \ref{fig:vdm_failures}.

The main exception is mIoU, where VDMs consistently outperform VLMs across all input conditions. This suggests that diffusion models are comparatively better at preserving coarse spatial occupancy and approximate object placement directly from visual observations. However, these spatially plausible extrapolation often deteriorate temporally, producing physically inconsistent motion or unstable long-horizon behavior despite maintaining visually reasonable foreground overlap.
\begin{wrapfigure}{r}{0.35\textwidth} 
  \centering
 \vspace{-3pt}
  \includegraphics[width=0.35\textwidth]{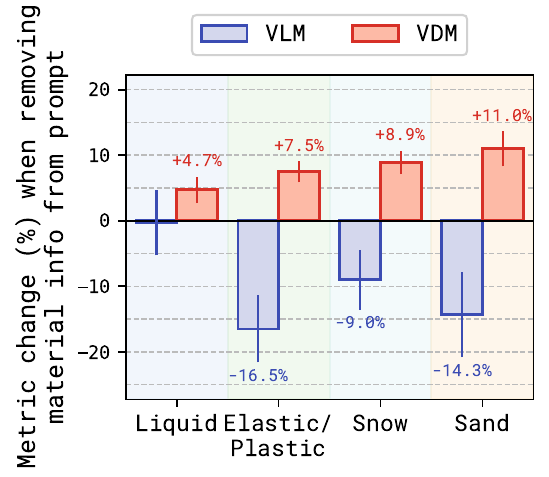}
  \caption{Performance change after removing material information, grouped by material family. 
  }
  \label{fig:material_sensitivity}
\end{wrapfigure}

\paragraph{How sensitive are models to structured scene information?} Figure \ref{fig:heatmap} shows that VLM performance is highly sensitive to the availability of structured scene information. Removing material or positional information substantially degrades performance, particularly for appearance fidelity (CTV) and motion consistency (W-MAE). In contrast, VDM performance changes comparatively little across input regimes, suggesting that current video diffusion models make limited use of additional structured physical information even when it is provided explicitly. Instead, they appear to rely primarily on short-term visual extrapolation.

Finally, both model families degrade substantially in the frames-only regime, indicating that long-horizon physical extrapolation remains difficult even when models observe several seconds of rollout history. However, the degradation is qualitatively different: VLM failures typically arise from incorrect physical implementation or unstable generated programs, whereas VDM failures are dominated by accumulated temporal drift, object instability, and incoherent motion over longer horizons. For additional qualitative results, see our \href{https://zzigak.github.io/mpmworlds/}{project website} and Appendix \ref{appendix:examples}.

\paragraph{Which materials are hardest to infer from visual information?}



Figure~\ref{fig:material_sensitivity} measures how performance changes when material information is removed from the structured prompt, relative to the full-configuration setting. For VLMs, removing material information has little effect on liquid scenes, but substantially degrades performance for elastic/plastic, snow, and sand scenes. This suggests that fluid-like behavior is either more readily inferred from the visual context or less sensitive to exact material parameters under our metrics, whereas coherent or history-dependent materials require more explicit material information.

In contrast, VDMs slightly improve when material information is removed across all material categories. This suggests that the VDM relies primarily on visual data and does not consistently benefit from textual material specifications. In some cases, additional structured material information may even act as noise for pixel-space extrapolation rather than improving physical consistency.

\begin{wrapfigure}{r}{0.37\textwidth} 
  \centering
  \vspace{-10pt} 
  \includegraphics[width=0.35\textwidth]{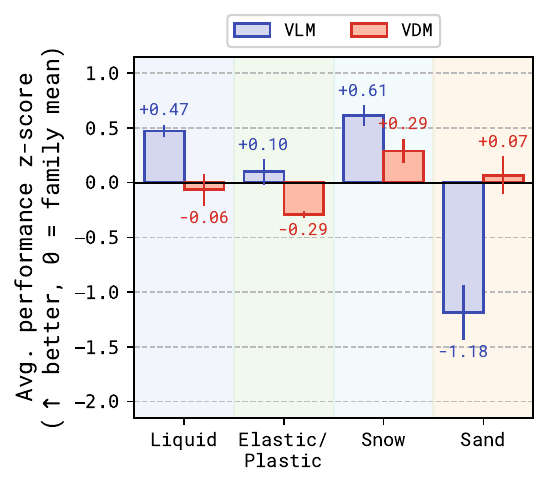}
  \caption{Average normalized performance by material family. Zero denotes each model-family mean across materials.}
  \label{fig:material_performance}
  \vspace{-10pt} 
\end{wrapfigure}

\paragraph{Which materials are most challenging for each model class?}

Figure~\ref{fig:material_performance} aggregates normalized metric performance across tasks for each material family. VLM-based approaches perform strongest on liquid and snow scenes, but struggle substantially on sand dynamics. This is consistent with the fact that sand is implementation wise among the most difficult constitutive models in our dataset, following the approach introduced in ~\citep{pradhanadrucker}. Although Drucker--Prager sand is also an elastoplastic material~\citep{li2026physics}, we separate it from elastic/plastic object scenes because it exhibits qualitatively different dynamics: granular flow, frictional yielding, and loss of coherent object shape. In contrast, our elastic/plastic category primarily contains coherent deformable bodies whose identity and geometry persist over time.

 In contrast, VDMs perform comparatively poorly on elastic and plastic scenes. Inspection of these rollouts suggests that the main failure mode is maintaining coherent object trajectories over time. Many elastic scenes contain bouncing or highly dynamic bodies whose motion requires temporally precise geometry and velocity propagation. Video diffusion models frequently produce drift, disappearance, or temporally inconsistent motion in these settings \ref{fig:vdm_failures}. Liquids, while visually complex, are often easier for VDMs because approximate flow structure is sufficient to maintain plausible appearance, even when fine-grained dynamics are inaccurate.
 

\paragraph{How do failure modes evolve over time?}

\begin{wrapfigure}{r}{0.35\textwidth} 
  \centering
  \vspace{-15pt} 
\includegraphics[width = 0.35\textwidth]{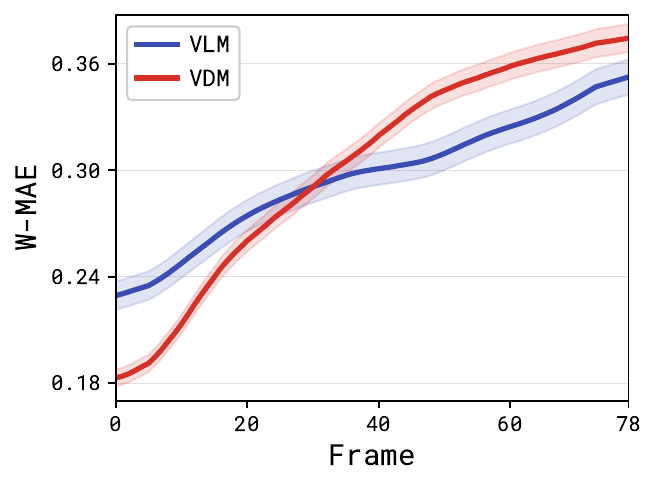}
  \caption{Moving-average W-MAE over continuation frames. VDM motion error grows faster with prediction horizon than VLM error.}
  \label{fig:temporal_wmae}
  \vspace{-15pt} 
\end{wrapfigure}


Figure~\ref{fig:temporal_wmae} shows the moving-average W-MAE across extrapolation frames, averaged over input conditions. The VDM's motion-activity error grows quickly after the prefix and remains consistently above the VLM curve, indicating that direct video prediction increasingly deviates from the ground-truth motion magnitude over time. The VLM also degrades with horizon, but more gradually, suggesting that executable simulation provides a more stable long-horizon dynamics prior.

We observe a similar pattern for collapse failures. Figure~\ref{fig:collapse_commulative} plots the cumulative fraction of samples that have collapsed by each extrapolation frame. VDM collapses occur early and continue accumulating throughout the rollout, reaching a substantially higher final collapse rate. In contrast, VLM rollouts exhibit far fewer collapses and mostly remain stable after the initial frames.

\begin{figure}[t]
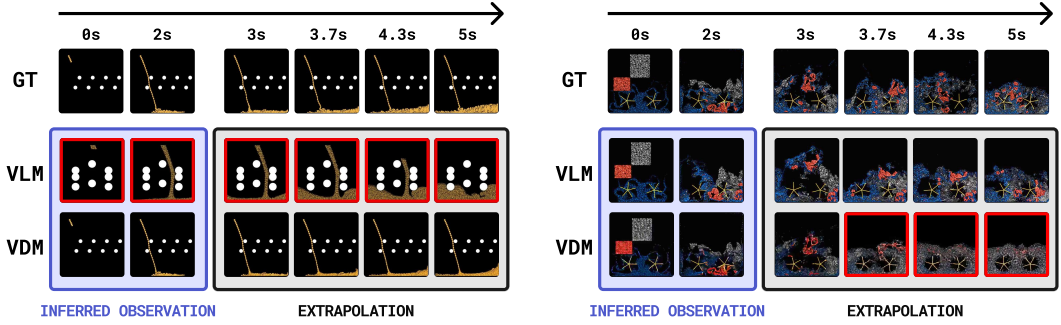

    \centering
    \begin{minipage}{0.48\textwidth}
        \centering
        \includegraphics[width=\linewidth]{images/position.pdf}
    \end{minipage}\hfill
    \begin{minipage}{0.48\textwidth}
        \centering
        \includegraphics[width=\linewidth]{images/color_coherence.pdf}
    \end{minipage}
    \caption{\textbf{VLM sensitivity to spatial inputs and compositional scaling.} \textbf{Left:} When explicit positional coordinates are withheld, the VLM struggles with visual state estimation, hallucinating an incorrect obstacle layout (red boxes) that alters the physical trajectory. The VDM, relying directly on pixels, preserves the geometry. \textbf{Right:} When provided with full configurations, the VLM scales seamlessly to compositional extremes (simultaneous liquid, elastic, and granular materials interacting with kinematic colliders). In contrast, the VDM suffers a catastrophic compositional breakdown, merging distinct materials into a single static average.}
    \label{fig:position_color}
    \vspace{-10pt}
\end{figure}

\paragraph{How well do models perform on modifications of previously observed scenes?}
Figure~\ref{fig:heatmap_modif} compares the fully held-out test split with the modification validation split, where models are evaluated on modified descendants of training scene families. Both model classes improve on the modification split, confirming that fully novel scene templates are harder than new variants of familiar scene structures.

The improvement is larger for VLMs, which gain across nearly all metrics and input regimes, especially in motion fidelity and temporal stability. VDMs also improve, but collapse and anomaly rates remain high, while mIoU remains their strongest metric. Thus, the VLM--VDM gap is not explained solely by memorization or scene novelty, suggesting that executable simulation generation benefits from familiar interaction structure, whereas pixel-space prediction remains better at geometric overlap but weaker at long-horizon dynamics. Figure \ref{fig:position_color} highlights these contrasting failure modes: VLMs struggle with visual state estimation when positional information is removed, while VDMs fail under compositional multi-material interactions despite preserving coarse geometry.



\paragraph{Can prefix quality predict extrapolation quality?} 

%

The VLM and VDM exhibit complementary strengths: VLM continuations are typically more physically consistent, while VDM continuations are often geometrically smoother. This motivates a simple question: can observable prefix quality predict which extrapolation to trust?

We test a lightweight prefix-quality gate. For each example, we evaluate how well the VLM reconstructs the visible prefix and use a learned threshold to decide whether to keep the VLM extrapolation or fall back to the VDM prediction. Full details are provided in Appendix \ref{appendix:prefix_gate}

 Table~\ref{tab:gate_top1} shows that this routing strategy consistently improves over either model alone, particularly for temporal and stability metrics such as W-MAE, anomaly rate, collapse score, and CTV, where the gate routes most samples to the VLM. In contrast, mIoU predominantly routes to the VDM, reflecting its stronger geometric consistency. Overall, these results suggest that prefix reconstruction quality is predictive of long-horizon extrapolation quality and that the two model classes capture complementary aspects of physical extrapolation.

\begin{table}[htbp]
\centering
\setlength{\tabcolsep}{4pt}
\caption{VLM vs VDM vs Threshold Gate (top-1) on the \textbf{test set}. {\color{myblue}\textbf{Blue}} = best, {\color{myred}red} = worst per metric. X\% of time the VLM output is selected by the gate.}
\label{tab:gate_top1}
\resizebox{\textwidth}{!}{%
\begin{tabular}{lrrrrrrrrrrrrrrr}
\toprule
\textbf{Input Type} & \multicolumn{3}{c}{\texttt{W-MAE}~$\downarrow$} & \multicolumn{3}{c}{\texttt{CTV}~$\downarrow$} & \multicolumn{3}{c}{\texttt{Anom.}~$\downarrow$} & \multicolumn{3}{c}{\texttt{Collapse}~$\downarrow$} & \multicolumn{3}{c}{\texttt{mIoU}~$\uparrow$} \\
\cmidrule(lr){1-1}\cmidrule(lr){2-4}\cmidrule(lr){5-7}\cmidrule(lr){8-10}\cmidrule(lr){11-13}\cmidrule(lr){14-16}
 & \textsc{vlm} & \textsc{vdm} & \textsc{gate} & \textsc{vlm} & \textsc{vdm} & \textsc{gate} & \textsc{vlm} & \textsc{vdm} & \textsc{gate} & \textsc{vlm} & \textsc{vdm} & \textsc{gate} & \textsc{vlm} & \textsc{vdm} & \textsc{gate} \\
\midrule
Frames + Full Config & 0.254 & 0.298 & {\color{myblue}\textbf{0.144}}~{\scriptsize\textit{88\%~LM}} & 0.188 & 0.251 & {\color{myblue}\textbf{0.132}}~{\scriptsize\textit{82\%~LM}} & 0.371 & 0.862 & 0.001~{\scriptsize\textit{97\%~LM}} & 0.163 & 0.366 & 0.126~{\scriptsize\textit{94\%~LM}} & 0.512 & 0.611 & {\color{myblue}\textbf{0.633}}~{\scriptsize\textit{23\%~LM}} \\
Frames + Config (No Materials) & 0.289 & 0.295 & 0.176~{\scriptsize\textit{80\%~LM}} & 0.190 & 0.274 & 0.139~{\scriptsize\textit{83\%~LM}} & 0.250 & {\color{myred}0.954} & 0.002~{\scriptsize\textit{100\%~LM}} & 0.125 & 0.399 & 0.073~{\scriptsize\textit{100\%~LM}} & 0.476 & 0.594 & 0.629~{\scriptsize\textit{20\%~LM}} \\
Frames + Config (No Positions) & 0.295 & 0.271 & 0.197~{\scriptsize\textit{95\%~LM}} & 0.267 & 0.249 & 0.193~{\scriptsize\textit{56\%~LM}} & 0.258 & 0.882 & {\color{myblue}\textbf{0.001}}~{\scriptsize\textit{100\%~LM}} & 0.114 & 0.338 & 0.063~{\scriptsize\textit{97\%~LM}} & 0.391 & 0.606 & 0.617~{\scriptsize\textit{2\%~LM}} \\
Frames Only & {\color{myred}0.349} & 0.345 & 0.221~{\scriptsize\textit{91\%~LM}} & {\color{myred}0.333} & 0.293 & 0.213~{\scriptsize\textit{39\%~LM}} & 0.276 & 0.711 & 0.001~{\scriptsize\textit{98\%~LM}} & 0.095 & {\color{myred}0.427} & {\color{myblue}\textbf{0.044}}~{\scriptsize\textit{95\%~LM}} & {\color{myred}0.316} & 0.578 & 0.617~{\scriptsize\textit{0\%~LM}} \\
\bottomrule
\end{tabular}
}
\end{table}

\section{Discussion}

We introduce a new dataset of physical simulations which help understand the strengths and weaknesses of code generation and video diffusion for physical inference and extrapolation.
Despite being limited to 2D physics, the simulations prove challenging for current approaches, owing to the rich variety of physical materials and their interactions instantiated in the data.
By feeding privileged information to the models we obtain insight on how to move forward: {VLMs benefit strongly from explicit scene information, suggesting they struggle to recover materials, positions, and parameters from vision alone.

\textbf{Toward Hybrid Architectures.} Our analysis of how models leverage side information demonstrates that code generation models (VLMs) excel at long-horizon stability but fail at visual state estimation, struggling to estimate geometry without explicit positional data. VDMs, in contrast, change little when given the same information, suggesting that their main bottleneck is not access to scene state, but using it to maintain coherent long-horizon dynamics. This points toward hybrid architectures that combine strong visual state estimation with explicit or structured dynamics representations. Indeed, we find that code generation and video diffusion can serve complementary roles for physical prediction, and our simple gating mechanism serves as an effective proof-of-concept for combining their strengths.

\paragraph{Limitations.}
Our dataset is limited to 2D MPM simulations and does not capture the full material or visual richness of the real world, including materials such as hair, cloth, or highly stiff bodies. Because all videos are generated by ground-truth simulation code, the benchmark may also favor code-generation approaches in ways that may not transfer to natural video. Future work should extend the benchmark to 3D scenes, real-world deformable dynamics, and natural-language side information, making models more useful for artists and everyday users. More broadly, realistic physical video generation may be misused to produce misleading simulations, making controlled evaluation and failure analysis important for responsible deployment.

\bibliographystyle{plainnat}
\bibliography{bib}

\newpage

\appendix
\clearpage
\section{Full test set evaluation results}
\begin{table*}[t]
\label{tab:full_test}

\centering
\setlength{\tabcolsep}{4.5pt}
\caption{Visual metrics for shared input regimes across VLM and VDM on \textbf{test holdout set} (top-1 per metric, mean\,$\pm$\,1\,SEM). {\color{myblue}\textbf{Blue bold}} = best, {\color{myred}red} = worst per metric.}
\label{tab:grouped_visual_metrics_top1}
\resizebox{\textwidth}{!}{%
\begin{tabular}{clrrrrrrr}
\toprule
& \textbf{Input Regime} & \texttt{WMAE}~$\downarrow$ & \texttt{CTV}~$\downarrow$ & \texttt{Anomaly}~$\downarrow$ & \texttt{Collapse}~$\downarrow$ & \texttt{mIoU}~$\uparrow$ & \texttt{SEMD}~$\downarrow$ & \texttt{MSE}~$\downarrow$ \\
\midrule
\multirow{4}{*}{\rotatebox{90}{\textit{VLM}}} & Frames + Full Config & \textbf{{\color{myblue}0.254 ± 0.026}} & \textbf{{\color{myblue}0.188 ± 0.017}} & 0.371 ± 0.042 & 0.163 ± 0.026 & 0.512 ± 0.021 & 0.064 ± 0.008 & 0.027 ± 0.002 \\
 & Frames + No Material Info & 0.290 ± 0.025 & 0.190 ± 0.014 & \textbf{{\color{myblue}0.250 ± 0.038}} & 0.125 ± 0.025 & 0.476 ± 0.020 & 0.066 ± 0.004 & 0.028 ± 0.002 \\
 & Frames + No Position Info & 0.295 ± 0.023 & 0.267 ± 0.014 & 0.258 ± 0.040 & 0.114 ± 0.024 & 0.391 ± 0.016 & 0.075 ± 0.004 & 0.036 ± 0.002 \\
 & Frames Only & {\color{myred}0.349 ± 0.025} & {\color{myred}0.333 ± 0.017} & 0.276 ± 0.040 & \textbf{{\color{myblue}0.095 ± 0.023}} & {\color{myred}0.316 ± 0.016} & {\color{myred}0.117 ± 0.015} & {\color{myred}0.037 ± 0.002} \\
\midrule
\multirow{4}{*}{\rotatebox{90}{\textit{VDM}}} & Frames + Full Config & 0.298 ± 0.018 & 0.251 ± 0.007 & 0.862 ± 0.028 & 0.366 ± 0.034 & \textbf{{\color{myblue}0.611 ± 0.015}} & \textbf{{\color{myblue}0.044 ± 0.002}} & \textbf{{\color{myblue}0.019 ± 0.001}} \\
 & Frames + No Material Info & 0.295 ± 0.018 & 0.274 ± 0.009 & {\color{myred}0.954 ± 0.017} & 0.399 ± 0.036 & 0.594 ± 0.015 & 0.045 ± 0.002 & \textbf{{\color{myblue}0.019 ± 0.001}} \\
 & Frames + No Position Info & 0.271 ± 0.017 & 0.249 ± 0.007 & 0.882 ± 0.026 & 0.338 ± 0.032 & 0.606 ± 0.015 & \textbf{{\color{myblue}0.044 ± 0.002}} & \textbf{{\color{myblue}0.019 ± 0.001}} \\
 & Frames Only & 0.345 ± 0.018 & 0.293 ± 0.011 & 0.711 ± 0.037 & {\color{myred}0.427 ± 0.035} & 0.578 ± 0.017 & 0.049 ± 0.004 & 0.020 ± 0.001 \\
\bottomrule
\end{tabular}
}
\end{table*}

\subsection{Example Results}
\label{appendix:examples}
Additional qualitative examples of model failure modes are provided in Figures \ref{fig:collapse_and_material} and \ref{fig:collapse_and_position}, illustrating object-collapse failures, material misidentification, and spatial reasoning errors across both model classes.

\begin{figure}[t]
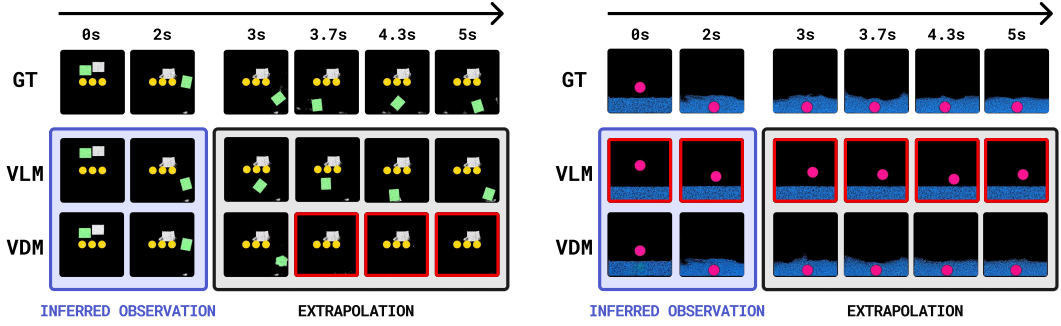

    \centering
    \begin{minipage}{0.48\textwidth}
        \centering
        \includegraphics[width=\linewidth]{images/object_collapse.pdf}
    \end{minipage}\hfill
    \begin{minipage}{0.48\textwidth}
        \centering
        \includegraphics[width=\linewidth]{images/material_fail.pdf}
    \end{minipage}
    \caption{\textbf{Failure modes in object permanence and material inference.} \textbf{Left:} In an elastoplastic scene, the VLM correctly maintains rigid object trajectories, while the VDM suffers from severe object collapse, causing the bouncing green block to disappear during extrapolation. \textbf{Right:} When explicit material properties are withheld from the input prompt, the VLM sometimes struggles with material inference from frames, hallucinating incorrect interaction dynamics (e.g., treating the liquid surface as a rigid boundary). The VDM, relying purely on pixel patterns, more accurately captures object materials.}
    \label{fig:collapse_and_material}
    \vspace{-10pt}
\end{figure}

\begin{figure}[t]
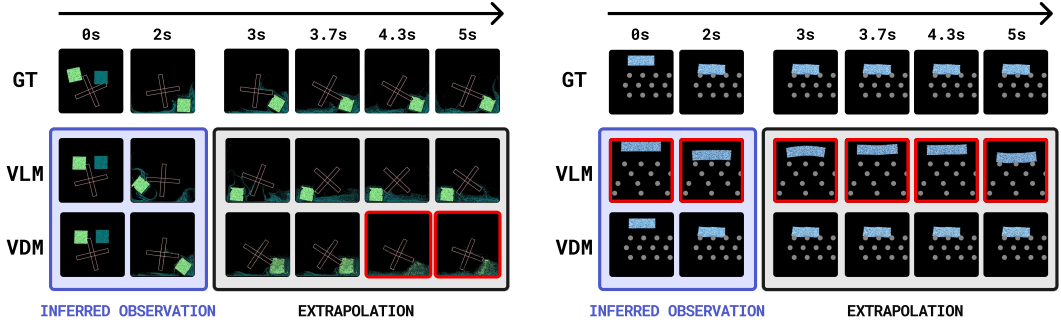

    \centering
    \begin{minipage}{0.48\textwidth}
        \centering
        \includegraphics[width=\linewidth]{images/object_collapse_2.pdf}
    \end{minipage}\hfill
    \begin{minipage}{0.48\textwidth}
        \centering
        \includegraphics[width=\linewidth]{images/minimal_positoin.pdf}
    \end{minipage}
    \caption{\textbf{Contrasting failure modes in complex dynamics and spatial reasoning.} \textbf{Left:} In a scene featuring high-energy multi-material interactions (a kinematic pinwheel, fluid, and a rigid block), the VLM accurately preserves object permanence and fluid volume. The VDM suffers from temporal object collapse, causing the rigid block to unphysically dissipate into the surrounding fluid during extrapolation (red boxes). \textbf{Right:} When explicit positional coordinates are limited, the VLM struggles with visual state estimation, hallucinating an incorrect geometric layout for the obstacle pegs (red boxes). The VDM, relying directly on pixel-space conditioning, successfully preserves the underlying spatial geometry.}
    \label{fig:collapse_and_position}
\end{figure}

\section{Evaluation Metrics}
\label{appendix:metrics}

\subsection{Mask Intersection-Over-Union}
For each frame, RGB images are thresholded into binary foreground masks and the Intersection-over-Union is computed frame-wise.  
$$\text{IoU}_t = \frac{|M_t^{\text{GT}} \cap M_t^{\text{pred}}|}{|M_t^{\text{GT}} \cup M_t^{\text{pred}}|},$$ with $\text{IoU}t = 1$ when both masks are empty. The final score is the mean over all video frames.

\subsection{Object Collapse Score (OCS)}
\label{appendix:ocs}

We measure prediction-internal temporal object coherence by checking whether objects generated early in the predicted rollout remain visible over time. 
We segment foreground components in an early frame using brightness thresholding and connected components, merge components with near-identical colors, and use the surviving components as predicted object/material regions. Each region defines an initial area $a_n^{(0)}$ and reference color $\mathbf{c}_n$.

At each later frame $t$, we estimate the area $a_n^{(t)}$ of region $n$ by counting foreground pixels whose chromaticity and brightness remain close to $\mathbf{c}_n$. The object-level collapse score is
$$
s_n^{(t)} = \max\left(0, 1 - \frac{a_n^{(t)}}{a_n^{(0)}}\right),
$$
and the per-video score is the worst collapse over all initialized regions and frames:
$$
\mathrm{OCS} = \max_{n,t>0} s_n^{(t)}.
$$
Lower values indicate that predicted objects or material regions remain stable; values near one indicate that at least one generated region nearly disappears during the predicted video. For example a value of 0.8 means on averaged over all test samples in that group, the single worst object in each generated video lost 80\% of its initial pixel count at some point during the predicted video.

\subsection{Temporal Anomaly Rate (RTSJ)}
\label{appendix:rtsj}
For each frame, we divide the image into coarse spatial regions and compute HSV color histograms within each region. Temporal jumps are measured by comparing histogram statistics across neighboring temporal windows, producing a localized spatiotemporal change signal. We then spatially aggregate this signal to obtain a jump magnitude $J_t$ for each timestep $t$.

For both the prediction and ground truth, we compute jump magnitudes $J_t^{\mathrm{pred}}$ and $J_t^{\mathrm{GT}}$. The excess jump is
\[
e_t=\max\left(J_t^{\mathrm{pred}}-J_t^{\mathrm{GT}},0\right).
\]
A prediction is flagged as anomalous if
\[
\mathrm{anomaly}=\mathbf{1}\left[\max_t e_t>\tau\right].
\]
The reported anomaly rate is the proportion of test samples flagged as anomalous.

\section{Modification test set values}

\begin{figure}[h]
\includegraphics[width = \textwidth]{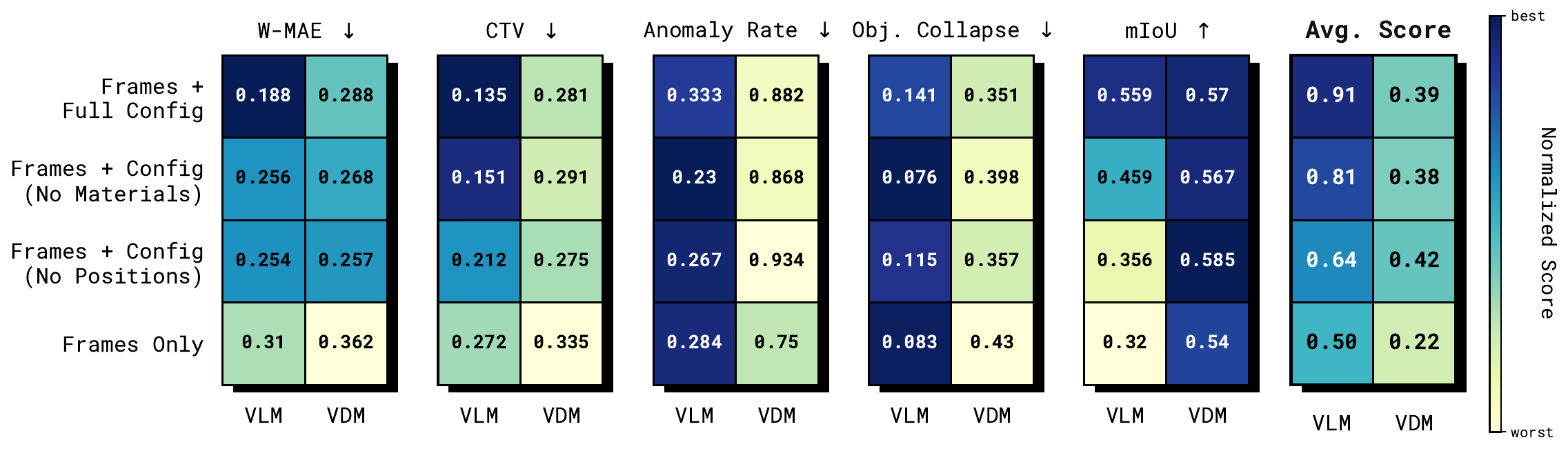}
\caption{ Comparison of VLM- and VDM-based extrapolation across input conditions and evaluation metrics. Values are averaged over the \textbf{modification test split}. Lower is better for W-MAE, CTV, anomaly rate, and object collapse, while higher is better for mIoU. The final column reports the mean normalized score across metrics for each model and input condition.}
\label{fig:heatmap_modif}
\end{figure}

\begin{table*}[t]
\centering
\setlength{\tabcolsep}{4.5pt}
\caption{Visual metrics for shared input regimes across VLM and VDM on \textbf{modification validation set} (top-1 per metric, mean\,$\pm$\,1\,SEM). {\color{myblue}\textbf{Blue bold}} = best, {\color{myred}red} = worst per metric.}
\label{tab:grouped_visual_metrics_top1_modif}
\resizebox{\textwidth}{!}{%
\begin{tabular}{clrrrrrrr}
\toprule
& \textbf{Input Regime} & \texttt{WMAE}~$\downarrow$ & \texttt{CTV}~$\downarrow$ & \texttt{Anomaly}~$\downarrow$ & \texttt{Collapse}~$\downarrow$ & \texttt{mIoU}~$\uparrow$ & \texttt{SEMD}~$\downarrow$ & \texttt{MSE}~$\downarrow$ \\
\midrule
\multirow{4}{*}{\rotatebox{90}{\textit{VLM}}} & Frames + Full Config & \textbf{{\color{myblue}0.188 ± 0.025}} & \textbf{{\color{myblue}0.135 ± 0.014}} & 0.333 ± 0.051 & 0.141 ± 0.028 & 0.559 ± 0.023 & 0.053 ± 0.005 & 0.018 ± 0.002 \\
 & Frames + No Material Info & 0.256 ± 0.031 & 0.151 ± 0.017 & \textbf{{\color{myblue}0.230 ± 0.045}} & \textbf{{\color{myblue}0.076 ± 0.022}} & 0.459 ± 0.028 & 0.081 ± 0.013 & 0.022 ± 0.002 \\
 & Frames + No Position Info & 0.254 ± 0.027 & 0.212 ± 0.017 & 0.267 ± 0.048 & 0.115 ± 0.029 & 0.356 ± 0.023 & 0.082 ± 0.012 & 0.026 ± 0.002 \\
 & Frames Only & 0.310 ± 0.030 & 0.272 ± 0.020 & 0.284 ± 0.048 & 0.083 ± 0.023 & {\color{myred}0.320 ± 0.022} & {\color{myred}0.090 ± 0.006} & {\color{myred}0.028 ± 0.002} \\
\midrule
\multirow{4}{*}{\rotatebox{90}{\textit{VDM}}} & Frames + Full Config & 0.288 ± 0.025 & 0.281 ± 0.018 & 0.882 ± 0.037 & 0.351 ± 0.043 & 0.570 ± 0.024 & 0.075 ± 0.017 & 0.015 ± 0.001 \\
 & Frames + No Material Info & 0.268 ± 0.025 & 0.291 ± 0.015 & 0.868 ± 0.039 & 0.398 ± 0.045 & 0.567 ± 0.023 & 0.067 ± 0.014 & 0.015 ± 0.001 \\
 & Frames + No Position Info & 0.257 ± 0.023 & 0.275 ± 0.016 & {\color{myred}0.934 ± 0.029} & 0.357 ± 0.043 & \textbf{{\color{myblue}0.585 ± 0.023}} & \textbf{{\color{myblue}0.050 ± 0.005}} & \textbf{{\color{myblue}0.014 ± 0.001}} \\
 & Frames Only & {\color{myred}0.362 ± 0.028} & {\color{myred}0.335 ± 0.021} & 0.750 ± 0.050 & {\color{myred}0.430 ± 0.044} & 0.540 ± 0.025 & 0.086 ± 0.019 & 0.015 ± 0.001 \\
\bottomrule
\end{tabular}
}
\end{table*}

\section{Prefix-Quality Threshold Gating}
\label{appendix:prefix_gate}

For each metric $m$ and task condition, we compute a prefix score
$
p_i^m = m(\hat{x}^{\mathrm{VLM}}_i, x_i)
$
between the predicted and ground-truth prefix, along with continuation scores
$
c_{i,\mathrm{VLM}}^m
$
and
$
c_{i,\mathrm{VDM}}^m
$
for the corresponding continuations.

We then sweep thresholds $\tau$ on a held-out validation set and select the threshold minimizing the average continuation error:
$$
\tau^{m*} =
\arg\min_{\tau}
\frac{1}{N}\sum_{i=1}^{N}
\begin{cases}
c_{i,\mathrm{VLM}}^m, & p_i^m < \tau, \\
c_{i,\mathrm{VDM}}^m, & \text{otherwise},
\end{cases}
$$
for lower-is-better metrics. For higher-is-better metrics such as mIoU, the inequality is reversed.

At inference time, we sample multiple VLM candidates, select the candidate with the best prefix score, and apply the learned threshold to decide whether to use the VLM or VDM continuation. Thresholds are learned independently for each metric and task condition.

\section{Temporal cumulative collapse rate}
\label{ref:cumulative_rate}
\begin{figure}[t]
\centering
\includegraphics[width = 0.35\textwidth]{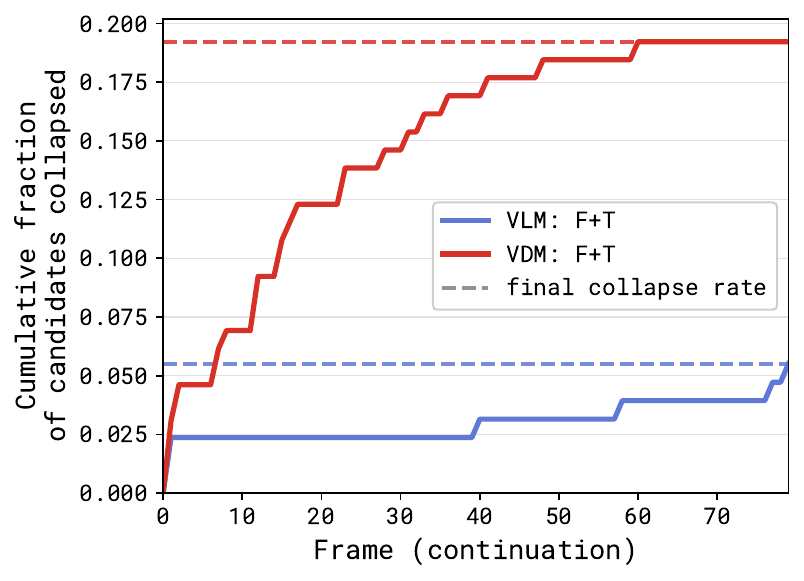}
\caption{Cumulative collapse rate over extrapolated frames. VDM rollouts collapse earlier and more often than VLM rollouts under the full-configuration setting.}
\label{fig:collapse_commulative}
\end{figure}

\section{Compute Resources.}
All VLM and VDM experiments were trained on a server with 4 NVIDIA RTX A6000 GPUs. Training each model required approximately 12 hours. Additional compute was used for dataset generation, rendering, and evaluation.

\end{document}